\pdfoutput=1
\documentclass[12pt]{iopart}

\renewcommand{\pt}{\ensuremath{p_{\rm t}}}
\newcommand{\kt}{\ensuremath{k_{\rm t}}}

\newcommand{\pp}{pp}
\newcommand{\PbPb}{Pb-Pb}

\usepackage{graphicx}

\begin{document}

\title[Jet background in heavy-ion collisions]{Jet reconstruction and
  jet background classification with the ALICE experiment in \PbPb\ collisions at the LHC}

\author{C Klein-B{\"o}sing{$^{1,2}$} for the ALICE Collaboration \footnote{For the full ALICE Collaboration author list and
acknowledgments, see Appendix "Collaborations" of this volume.}}

\address{{$^1$} Institut f{\"u}r Kernphysik -  M{\"u}nster, Germany}
\address{{$^2$} ExtreMe Matter Institute, GSI - Darmstadt, Germany}
\ead{Christian.Klein-Boesing@uni-muenster.de}
\begin{abstract}
  For a quantitative interpretation of reconstructed jet
  properties in heavy-ion collisions it is paramount to characterize
  the contribution from the underlying event and the
  influence of background fluctuations on the jet signal. 
  In addition to the pure number
  fluctuations, region-to-region correlated background within one
  event can enhance or deplete locally the level of background and
  modify the jet energy. We show a first detailed assessment of
  background effects using different probes embedded into heavy-ion
  data and quantify their influence on the reconstructed jet spectrum.

\end{abstract}
 
\section{Introduction}

The quantification of the effect of parton energy loss, known as jet
quenching, is one of the major goals of jet and high \pt\ measurements
in heavy-ion collisions. The aim of jet reconstruction is to gain more
direct access to the orignal parton properties and the modification of
their fragmentation process in heavy-ion collisions than single particle
measurements \cite{Salgado:2002ws,Alessandro:2006yt}.
Already the first measurements of
reconstructed jets in heavy-ion collisions at the LHC showed a
striking imbalance between back-to-back dijets \cite{Aad:2010bu}
pointing to a significant partonic energy loss in the medium. However,
a quantitative interpretation of any jet result requires a precise
knowledge of the background-induced fluctuations of the measured jet signal,
which can distort the jet balance even in the absence of any other
medium effects \cite{Cacciari:2011tm}.

\section{Jet Reconstruction and Background Subtraction}

\begin{figure}
\begin{center}
   \unitlength1cm
    \begin{picture}(16,5.3)
   \put(0,0){\includegraphics[height=5.3cm]{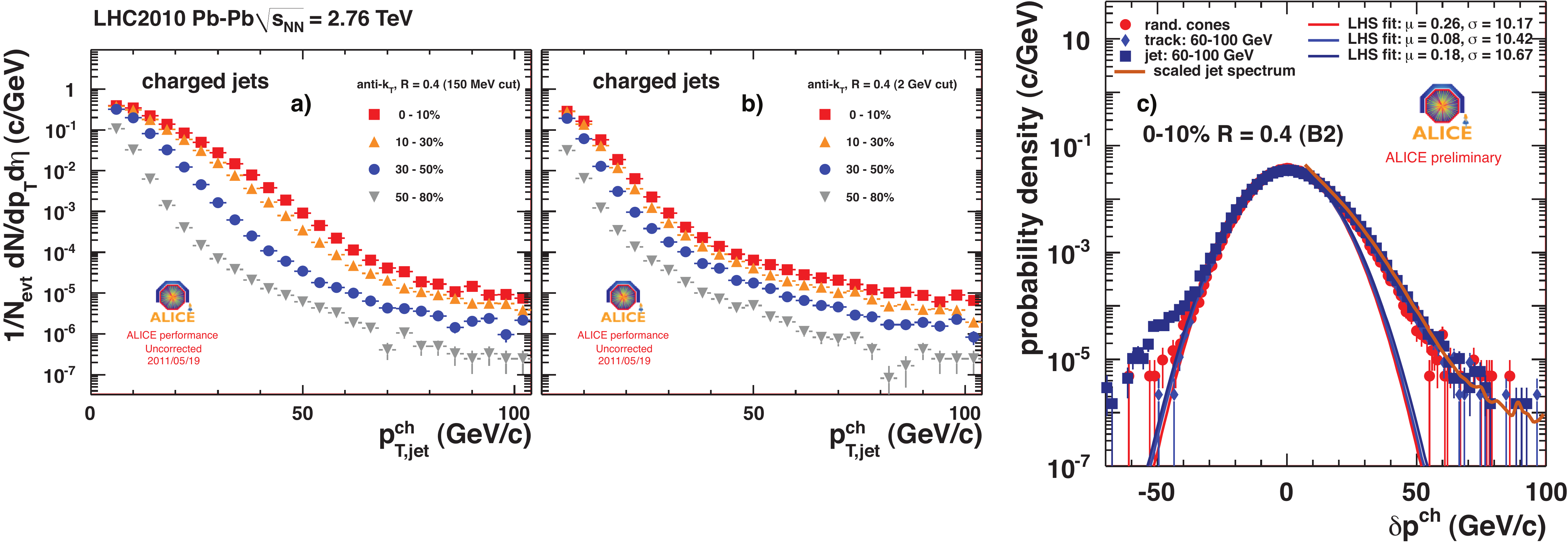}}
\end{picture}
\vspace{-0.8cm}
\end{center}
  \caption{
    \label{fig:jetspec}
    Reconstructed raw jet spectra for different centralities using the anti-\kt\ algorithm after background
    subtraction: a) track $\pt > 150$ MeV/$c$
    b) track $\pt >  2$ GeV/$c$. c) $\delta{\pt}$ distribution of charged particles in  the 10\% most central
    \PbPb\ events for different probes together with scaled jet
    spectrum ( {\pt} cut of 150 MeV/$c$).}
\end{figure}

The data presented here were collected by the ALICE experiment \cite{Aamodt:2008zz} in the first \PbPb\
run of the LHC in the fall of 2010 with an energy of $\sqrt{s_{\rm NN}} =
2.76$~TeV. Since the Electromagnetic Calorimeter (EMCal) has only been fully
installed since the beginning of 2011, jets from the first \PbPb\ and \pp\
collisions in ALICE are reconstructed based on charged particles
only. For this we use tracks reconstructed in the
Time-Projection-Chamber (TPC) together with vertexing information from the
Inner Tracking System (ITS). This ensures maximum azimuthal angle ($\phi$)
uniformity of reconstructed tracks with transverse momenta down to $\pt = 150$~MeV/$c$.

We employ a variety of jet finders, which exhibit different
sensitivities to the presence of large backgrounds in heavy-ion
collisions and show good agreement above $\pt = 20$~GeV/$c$
in \pp-collisions \cite{KleinBoesing:2011vx}; cone algorithms (UA1 and SISCone), as well as the
sequential recombination algorithms from the FastJet package ({\kt}
and {anti-\kt} \cite{Cacciari:2005hq}), all with a distance/radius
parameter of 0.4. Here, the
$\kt$ algorithm is used to estimate the background density $\rho$ on
an event-by-event basis by calculating the median $\pt$/area of
reconstructed $\kt$-clusters in $\eta < 0.5$ after removing the two
leading clusters (see also \cite{Cacciari:2007fd}). For the present
study we focus on jets that are reconstructed using the anti-$\kt$
algorithm and corrected for the background density in each event using
the jet area $A$ with $\pt^{\rm jet} = \pt^{\rm jet,rec} - \rho \cdot
A$.

The resulting raw jet spectra are shown in figure~\ref{fig:jetspec}a)
and b) for different values of the minimum track $\pt$. The difference
in the shape for central reactions and the low $\pt$ track cut is
clearly visible, showing the dominance of background fluctuations over a wide
$\pt$-range.

\section{Background Fluctuations}

\begin{figure}
\begin{center}
   \unitlength1cm
    \begin{picture}(16,5.3)
     \put(0,0){\includegraphics[height=5.3cm]{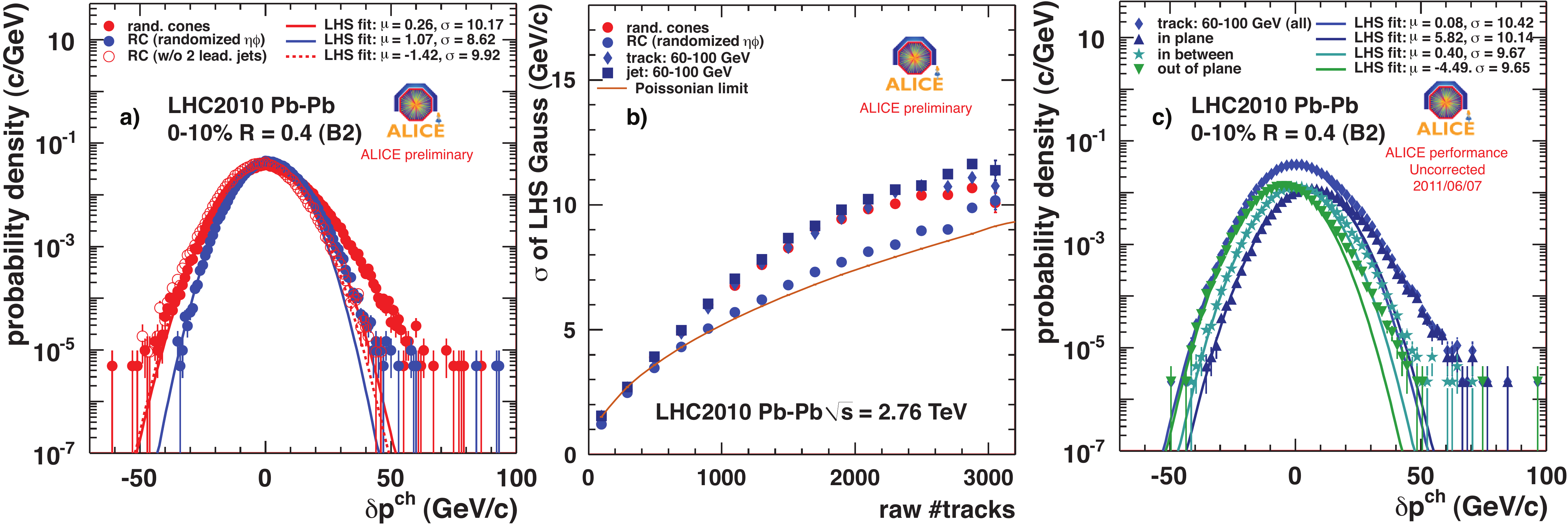}}
\end{picture}
\vspace{-0.8cm}
\end{center}
  \caption{  $\delta{\pt}$ distribution of charged particles in  the 10\% most central
    \PbPb\ events  for a) random cones (RC) on full
    events, requiring a minimum distance to the two leading jets, and
    on randomized events c) for single tracks for
    different orientations to the event plane. b) Evolution of background
    fluctuations with input multiplicity and compared to the Poissonian
    limit derived from the measured track \pt\ spectrum. 
    \label{fig:deltaptAll}
}
\end{figure}
Background fluctuations are determined using different probes of the
measured \PbPb\ collisions: random cones ($R = 0.4$) placed into the jet
acceptance, where the \pt\ of all tracks in the cone is summed,
embedding of single high \pt\ tracks, and embedding of full \pp\ jet events (real or
from full detector simulation). In case of embedding the standard
anti-\kt\ algorithm with $R = 0.4$ is used to cluster the event and
the reconstructed jets are matched to the embedded probe, by either
finding the single track in it, or 50\% of the \pp\ jet momentum (in
the case of jet embedding).
The residuals of the background subtraction are then given  by:
\begin{equation}
 \delta{\pt} = \pt^{\rm rec} - A \cdot \rho - \pt^{\rm probe}
\end{equation}
(where $\pt^{\rm probe}$ is zero in the case of random cones). The
resulting distributions are shown in figure~\ref{fig:jetspec}c). The
distribution is peaked around zero, illustrating the validity of the
background subtraction. We iteratively fit the left-hand-side of each
distribution with a Gaussian to obtain a measure for the width of the
distribution. The $\sigma$ of the Gaussian fit provides the lower
limit on the total fluctuations and is larger than 10 GeV/$c$ for all
methods. The deviation from the Gaussian shape on the right-hand-side
of the distribution is apparent, and is also reflected in the RMS of
the full distributions which ranges from 11.4 to 12 GeV/$c$. It can be
explained mainly by the presence of jets in the \PbPb\ events used for
embedding, as seen by the shape of the scaled reconstructed jet
spectrum.  It is also seen in figure~\ref{fig:deltaptAll}a), where the
$\delta{\pt}$ distribution from random cones on full events is
compared to random cones with a minimum distance to the leading jets
and in events where the tracks have been randomized in $\eta$ and
$\phi$. In the latter two cases the tail attributed to jets is much
less pronounced and a Gaussian shape is almost recovered.  It has to
be noted that the distributions due to purely statistical fluctuations
are not expected to exactly follow a Gaussian shape, but instead are
better described by a $\Gamma$-function \cite{Tannenbaum:2001gs}, since
$\delta{\pt}$ is similar to a measurement of $\left<\pt\right>$ fluctuations in a
limited region of phase space. The removal of the jet contribution
does not affect the left side of the distribution, it only becomes
narrower if all residual correlations are destroyed by randomizing the
event, this points to correlated region-to-region fluctuations in
addition to the purely statistical.

 The contribution of statistical fluctuations to the overall
 fluctuations of the underlying event within a typical jet cone can be
 seen in figure~\ref{fig:deltaptAll}b), where the Poissonian limit
 estimated via:
\begin{equation}
\mathrm{RMS}(\delta{\pt}) = \sqrt{N_{\rm A}} 
\cdot \sqrt{\langle \pt \rangle ^2 + \mathrm{RMS}(\pt)^2}
\end{equation}
is compared to the measured fluctuations for different input
multiplicities. Here, $N_{\rm A}$ is the total uncorrected input
multiplicity scaled to the jet area. The values are taken from the
uncorrected track \pt\ spectrum used for jet finding.  For ideal track
detection the statistical fluctuations in a typical jet cone would
increase by 8\%,
using the values from the efficiency and acceptance corrected track \pt\
spectrum. The $\sqrt{N}$ increase is clearly seen in all cases, but
only for randomized events the limit is approached, with large
differences especially at intermediate multiplicities.

One natural source of region-to-region differences is the presence of
collective effects (flow) in the underlying event. This can be
visualized when dividing the embedded probes into different bins,
depending on their orientation to the event plane. In
figure~\ref{fig:deltaptAll}c) it can be clearly seen that for probes embedded
out of plane the background is overestimated by almost 5 GeV/$c$, for
in plane the effect is opposite. Since within the jet we sum over all
particle $\pt$ the effect scales with $\sum v_2(\pt) \pt$ and thus is
still sizable in central events. The separation of collective effects
from jet-induced region-to-region fluctuations will be a challenge for
any jet measurement at lower \pt\ and with low momentum cut-off, but
is essential for the understanding of the modification of jet
fragmentation at low track $\pt$ and its path-length dependence.

\section{Summary}

We have presented the first detailed study of background fluctuations for
jet reconstruction in \PbPb\ collisions at the LHC, where we observe
for charged particle jet reconstruction and a low \pt\ cut-off of 150
MeV/$c$ jet background fluctuations larger than 10 GeV/$c$ in central
collisions and for a typical distance parameter/radius of 0.4. The
fluctuations show a significant increase compared to purely
statistical fluctuations, which is caused by correlated region-to-region fluctuations. In particular the presence of collective effects
(flow) can induce shifts in the background subtracted jet $\pt$ of up
to 5 GeV/$c$.

\end{document}